%% file: betalambda_EW_jhep_v4.tex
\documentclass[11pt,a4paper]{article}
\usepackage{jheppub}

\input{macros}

\allowdisplaybreaks
\newcommand{\ice}[1]{\relax}

\title{$\beta$-function for the Higgs self-interaction in the Standard Model at three-loop level}
\author[a]{K. G. Chetyrkin}
\author[a]{and M. F. Zoller}
\affiliation[a]{Institut f\"ur Theoretische Teilchenphysik, Karlsruhe
  Institute of Technology (KIT), \mbox{D-76128 Karlsruhe, Germany}}
\emailAdd{k.chet@kit.edu}
\emailAdd{max.zoller@kit.edu}

\abstract{We analytically compute the QCD, electroweak, Higgs and third generation Yukawa contributions to the
$\beta$-function for the Higgs self-coupling as well as for the Higgs mass parameter
in the unbroken phase of the Standard Model at three-loop level.}

\arxivnumber{}
\subheader{TTP13-008\\SFB/CPP-13-19}

\begin{document}
\maketitle

\section{Introduction}
Renormalization Group functions, i.e. $\beta$-functions and anomalous dimensions, are of great importance
in quantum field theory. Recently most of these functions have been computed at three-loop accuracy in the Standard Model (SM).
The results for the gauge couplings have first been derived in \cite{PhysRevLett.108.151602,Mihaila:2012pz}
and have been confirmed independently in \cite{Bednyakov:2012rb}. For the top-Yukawa coupling and the Higgs 
self-interaction $\lambda$ the QCD, top-Yukawa and Higgs contributions have been derived in \cite{Chetyrkin:2012rz}. 
The result for the top-Yukawa $\beta$-function
has been extended to include the electroweak and all third generation Yukawa couplings in \cite{Bednyakov:2012en} where also the
$\beta$-functions for bottom and $\tau$-Yukawa have been presented.
The one-loop and two-loop results for all SM couplings have been known for a long time
\cite{PhysRevLett.30.1343,PhysRevLett.30.1346,Jones1974531,Tarasov:1976ef,PhysRevLett.33.244,
Egorian:1978zx,PhysRevD.25.581,Fischler:1981is,Fischler1982385,Jack1985472,
Machacek198383,Machacek1984221,Machacek198570,2loopbetayukawa,Ford:1992pn} as have been partial three-loop
results \cite{Curtright:1979mg,Jones:1980fx,Tarasov:1980au,3loopbetaqcd,Steinhauser:1998cm,Pickering:2001aq}.
Four-loop $\beta$-functions are available for QCD \cite{4loopbetaqcd,Czakon:2004bu} and the purely scalar part of the SM \cite{Brezin:1974xi,Brezin:1973,Kazakov_betalambda}.
In this paper we present the extension of our result for the Higgs self-interaction and 
the anomalous dimension of the Higgs mass parameter to include the electroweak and all third generation Yukawa couplings
as well.
Especially the $\beta$-function for the Higgs self-coupling is
interesting because of its close connection to the question of vacuum
stability in the Standard Model. It has been shown that the stability
of the SM vacuum up to some energy scale $\Lambda$ is approximately
equivalent to the requirement that the running coupling
$\lambda(\mu)>0$ for $\mu \leq \Lambda$
\cite{Cabibbo:1979ay,Ford:1992mv,Altarelli1994141}. Many analyses of
this question have been performed
\cite{Bezrukov:2009db,Holthausen:2011aa,EliasMiro:2011aa,Xing:2011aa,
Bezrukov:2012sa,Degrassi:2012ry,Chetyrkin:2012rz,Masina:2012tz} during the last years. The main uncertainty
stems from the experimental error on the top mass followed by the uncertainty in $\als$.
But future linear colliders could greatly reduce these uncertainties
and a possible Higgs mass of about $126$ GeV
\cite{ATLAS:2012ae,Chatrchyan:2012tx} allows for both scenarios, a
stable and an unstable (or to be more precise metastable) SM vacuum,
within the present experimental and theoretical errors. For this
reason we think that the present work will reduce the theoretical
uncertainty connected to the running of $\lambda$ even further than our
previous calculation \cite{Chetyrkin:2012rz}. 

In the following section the setup and the technical details of the calculation are discussed.
After that we present our results and some numerics in order to determine the significance of the new terms.

\section{Calculation}
The gauge group of the SM is an $\text{SU}_{\sss{C}}(3)\times \text{SU}(2)\times
\text{U}_{\sss{Y}}(1)$ which is spontaneously broken to an
$\text{SU}_{\sss{C}}(3)\times \text{U}_{\sss{Q}}(1)$ at the Fermi scale. 
Our calculation is performed in the unbroken phase of the SM which is justified
by the fact that the UV behaviour and therefore the renormalization constants for fields and vertices do not depend on
masses \ice{and external momenta} in the $\overline{\text{MS}}$-scheme \cite{Collins:1974da}.

The Lagrangian of the SM can be decomposed into the following pieces:
\be
\ssL=\ssL_{\sss{QCD}}+\ssL_{\sss{EW}}+\ssL_{\sss{Yukawa}}+\ssL_{\sss{\Phi}}.
\ee
The QCD and electroweak (EW) part are implemented in the usual way with the gauge fields $A^a_\mu$ ($\text{SU}_{\sss{C}}(3)$,
$a=1,\ldots, 8$),
$W^a_\mu$ (SU$(2)$, $a=1,2,3$) and $B_\mu$ ($\text{U}_{\sss{Y}}(1)$). These appear in the covariant derivative
\be
D^\mu=\p^\mu-i g_1 Y_f B^\mu-i \f{g_2}{2} \sigma^a W^{a\,\mu}-i\gs T^a A^{a\,\mu} \label{covderiv}
\ee
with the Pauli matrices $\sigma^a$ and the hypercharge $Y_f$ of the field $f$ on which the covariant derivative acts.  
In the Yukawa part we neglect the first two generations and the mixing of generations. 
The W-fermion-vertices are taken to be diagonal in the generations as well, i.e. we set the CKM matrix to the unit matrix.
Light quarks and leptons are present in the QCD and electroweak sector however. This leads to the Lagrangian 
\be
\begin{split}
\ssL_{\sss{Yukawa}}=&-\yt \left\{\bar{t}_{\sss{R}}\Phi^{\dagger\,c}Q_{\sss{L}}+\bar{Q}_{\sss{L}}\Phi^c t_{\sss{R}}\right\}
-\yb \left\{\bar{b}_{\sss{R}}\Phi^{\dagger}Q_{\sss{L}}+\bar{Q}_{\sss{L}}\Phi t_{\sss{R}}\right\}\\
&-\ytau \left\{\bar{\tau}_{\sss{R}}\Phi^{\dagger}L_{\sss{L}}+\bar{L}_{\sss{L}}\Phi t_{\sss{R}}\right\}
\end{split} 
\label{LYuk} 
\ee
for the Yukawa sector.
The complex scalar field $\Phi$ and the left-handed quarks and leptons are doublets 
under SU$(2)$:
\be
\Phi=\vv{\Phi_1}{\Phi_2},\qquad \Phi^c=i\sigma^2\Phi^{*},\qquad Q_{\sss{L}}
=\vv{t}{b}_{\sss{L}},\qquad L_{\sss{L}}=\vv{\nu_\tau}{\tau}_{\sss{L}}
{}.
\ee
The indices L and R indicate the left- and right-handed part of the fields as obtained by the projectors
\be 
P_{\sss{L}}=\f{1}{2}\lb 1-\gamma_5\rb \qquad P_{\sss{R}}=\f{1}{2}\lb 1+\gamma_5\rb
{}.
\ee
Finally, we have the Higgs sector with
\be
\begin{split}
\ssL_{\sss{\Phi}}=(D_\mu \Phi)^\dagger (D^\mu \Phi)-m^2\Phi^\dagger\Phi-\lambda \lb\Phi^\dagger\Phi\rb^2.
\end{split} 
\label{LPhi} 
\ee  

For every field and vertex a counterterm is introduced and the corresponding renormalization constant is calculated
order by order in perturbation theory.
The left-handed and right-handed parts of fermion fields, quark-gluon-vertices and fermion-B-vertices are renormalized
with different counterterms.
The renormalization constant for the gauge, Yukawa and Higgs couplings can be obtained in different ways, e.g.
\be 
Z_{g_1}=\f{Z^{(\tau\tau B)}_{1,L}}{Z^{(2\tau)}_{2,L}\sqrt{Z^{(2B)}_3}}=
\f{Z^{(tt B)}_{1,R}}{Z^{(2t)}_{2,R}\sqrt{Z^{(2 B)}_3}}=\ldots \ee
where $Z_1^{f_1 \ldots f_n}$ is the renormalization constant for the vertex of the (renormalized) fields $f_1,\ldots,f_n$,
$Z^{(2f)}_{2,L/R}$ the field strength renormalization constant for the left-handed (L)/ right-handed (R) part of the fermion field $f$
and $Z^{(2 g)}_3$ the field strength renormalization constant for the gauge field $g$. 

Likewise, the renormalization constant for Yukawa couplings can be computed from the renormalization constant for
any vertex proportional to this coupling and the renormalization constants for the external legs of this vertex, e.g.
\be 
Z_{\yt}=\f{Z^{(tt\Phi)}_1}{\sqrt{Z^{(2t)}_{2,L}Z^{(2t)}_{2,R}Z_2^{(2\Phi)}}}
{},
\ee
where $Z_2^{(2\Phi)}$ is the field strength renormalization constant for the scalar doublet.
All renormalization constants are defined in a minimal way as
\be Z=1+\delta Z{}, \ee
with $\delta Z$ containing only poles in the regulating parameter  $\eps = (4-D)/2$ where
$D$ is the engineering space-time dimension.
The Higgs self-coupling $\lambda$ is renormalized with a counterterm $\delta Z_\lambda$ that is not proportional to $\lambda$ but
also has terms proportional to four Yukawa couplings. Consequently, this is a feature of the corresponding $\beta$-function as well. We find
\be \begin{split}
     -\lambda_B\left(\Phi_B^\dagger\Phi_B\right)^2=&(-\lambda+\delta Z_1^{(4\Phi)})\left(\Phi^\dagger\Phi\right)^2\\
     \Rightarrow -(\lambda+\delta Z_\lambda) \left(Z_2^{(2\Phi)}\right)^2=&(-\lambda+\delta Z_1^{(4\Phi)})\\
     \Rightarrow \delta Z_\lambda=& \left(Z_2^{(2\Phi)}\right)^{-2} (\lambda-\delta Z_1^{(4\Phi)})-\lambda{}\,,
    \end{split}
\ee
with the counterterm $\delta Z_1^{(4\Phi)}$ for the four-$\Phi$-vertices and the index $B$ marking bare quantities.
The $\beta$-function for any coupling X is defined as
\be
\beta_{\sss{X}}=\mu^2\f{d X}{d \mu^2}    =\sum \limits_{n=1}^{\infty} \f{1}{(16\pi^2)^{n}}\,\beta_{\sss{X}}^{(n)}
\ee
and is given as a power series in all considered couplings of the SM, i.e. $\gs, g_2, g_1, \yt, \yb, \ytau$ and $\lambda$. 
The mass parameter $m^2$ of the scalar field is neglected in the calculation as it has no influence on the UV behaviour
of the couplings in the $\overline{\text{MS}}$-scheme.

The $\beta$-function (or anomalous dimension) describing the running of this mass parameter $m^2$ in eq.~(\ref{LPhi}) 
can be computed from the renormalization constant of
the local operator $O_{2\Phi}:=\Phi^\dagger \Phi$. An insertion of $O_{2\Phi}$ into a Green's function, 
e.g. with two external $\Phi$-fields,
is renormalized as $[O_{2\Phi}]=Z_{\Phi^2}O_{2\Phi}$ where $[O_{2\Phi}]$ is the corresponding finite operator.
From $[O_{2\Phi}]=Z_{m^2}O^{\text{bare}}_{2\Phi}$ and $O^{\text{bare}}_{2\Phi}=Z_2^{(2\Phi)}O_{2\Phi}$ it follows that
\be Z_{m^2}=\lb Z_2^{(2\Phi)}\rb^{-1} Z_{\Phi^2}
{}. 
\ee 
For this project we need the renormalization constants $\delta Z_1^{(4\Phi)}$, $Z_{\Phi^2}$ and $Z_2^{(2\Phi)}$ at three-loop
level and all other counterterms at two-loop accuracy or less. 
We perform our calculation in a general $R_\xi$-gauge with different gauge parameters $\xi_1$, $\xi_2$ and $\xi$ for the gauge
fields $B$, $W$ and $A$. The $\beta$-functions for all couplings are independent of the gauge parameters
which serves as an important check for the result.
In order to compute all counterterms up to two-loop level and the one and two-loop diagrams with counterterm insertions
contributing to the three-loop result it has been convenient to use a setup where different isospin configurations of a field
and different fermion generations (in our case the distinction between third generation and light is enough)
are implemented as separate fields as many counterterms depend on those. Additionally, the $\text{U}_{\sss{Y}}(1)$ hypercharge depends
on the isospin and differs for the left-handed and right-handed part of the field. So we use the set of fields
$$
\underbrace{t,b,u,d}_{\text{quarks}},\underbrace{e^{-},\nu_e,\tau,\nu_\tau}_{\text{leptons}},
\underbrace{A^{a\mu},B^\mu,W^{1\mu},W^{2\mu},W^{3\mu}}_{\text{gauge bosons}},\underbrace{c^a,c_W^1,c_W^2,c_W^3}_{\text{ghosts}},
\Phi_1,\Phi_2
$$
and their anti-fields.
The fermion fields have to be split up in a left-handed and a right-handed part during the calculation. 
The price we pay, however, is that many diagrams are produced which look the same in momentum space if no counterterms.
are inserted. For the 1PI process
with four external $\Phi$-legs $\sim 2.3 \times 10^6$ diagrams are generated at three-loop level. In order to reduce this number and
because we do not need three-loop diagrams with counterterm insertions anyway we have chosen a second smaller
set of fields for this part of the calculation
$$
\underbrace{q_{i}}_{\text{quarks}},\underbrace{l_{i}}_{\text{leptons}},
\underbrace{A^{a\mu},B^\mu,W^{a\mu}}_{\text{gauge bosons}},\underbrace{c^a,c_W^a}_{\text{ghosts}}, \Phi_i
{}.
$$
Here the index $i$ marks the isospin of $\Phi$ and the left-handed fermions. For the right-handed fermions
$i=1,2$ is just a label to mark the flavour, e.g. $q_{1,R}=t_R, q_{2,R}=b_R$.
Fermion loops without Yukawa interactions are multiplied by the number of generations $\NGen$ in order to include the light
fermions. The indices of external particles can be explicitly chosen. Using this setup only $573692$ diagrams
are produced for the four-$\Phi_1$-process at three loops.
The computation of the $\text{SU}(2)\times \text{U}_{\sss{Y}}(1)$ group factors has been implemented with Mathematica
using labels for the left-handed and right-handed part at each quark-Yukawa-vertex
and $B$-fermion-vertex as well as for the three different structures in the four-$W$-vertex. With the help of these labels
we can completely factorize the $\text{SU}(2)\times \text{U}_{\sss{Y}}(1)$ part from the momentum space diagram.
The QCD colour factors have been calculated with the FORM package COLOR \cite{COLOR}. 
All Feynman diagrams have been automatically generated with QGRAF \cite{QGRAF}.

In order to check our setup we have computed all Yukawa and gauge coupling renormalization constants at two-loop level
from at least two different vertices with the first set of fields and compared the result to the literature.
The same has been done for the renormalization constants of the gauge, ghost and scalar fields.
We also explicitly checked that we get the same renormalization constants for $\Phi_1$ and $\Phi_2$ as well as
for the left-handed fermion flavours of the same generation at two loops. Another check has been the finiteness of the three-$B$-vertex
up to two loops.\\
The renormalization constant for $\Phi$ has been computed with both sets of fields at three-loop level which yields
the same result as in \cite{Mihaila:2012pz}. The renormalization constants for the $W$, $c_W$, $\Phi_1$ and $\Phi_2$ fields
as well as for the \mbox{$W$-$\bar{c}_W$-$c_W$-vertex}, the \mbox{$O_{2\Phi}$-$\Phi_1$-$\Phi_1$-vertex}, the 
\mbox{$O_{2\Phi}$-$\Phi_1$-$\Phi_1$-vertex}
and the \mbox{$4$-$\Phi_1$-vertex} have been computed with both sets of fields up to two loops with the same result.

As explained in detail in \cite{Chetyrkin:2012rz} 
some diagrams with four external $\Phi$-fields where two
external momenta are set to zero suffer from IR divergences which
mix with the UV ones in dimensional regularization. We therefore use the same method as in \cite{Chetyrkin:2012rz}
and introduce the same auxiliary mass parameter $M^2$ in every propagator denominator. Subdivergences $\propto M^2$ 
are canceled by counterterms
\be \begin{split}
\f{M^2}{2}\delta\!Z_{\sss{M^2}}^{(2g)}\,A_\mu^a A^{a\,\mu},\;
\f{M^2}{2}\delta\!Z_{\sss{M^2}}^{(2W)}\,W_\mu^a W^{a\,\mu},\; 
\f{M^2}{2}\delta\!Z_{\sss{M^2}}^{(2B)}\,B_\mu B^{\mu}\;
\text{and} \;
\f{M^2}{2}\delta\!Z_{\sss{M^2}}^{(2\Phi)}\, \Phi^\dagger\Phi{}.
\end{split} \ee
Counterterms $\propto M$ that would arise for fermions cannot appear because there are no $M$ in the numerators
of propagators. Ghost mass terms $\f{M^2}{2}\delta\!Z_{\sss{M^2}}^{(2c)}\,\bar{c}^a c^a$ for the SU$(3)$ and SU$(2)$ ghosts
do not appear because of the momentum dependence of the ghost-gauge boson-vertex.
The remaining divergences are the mass-independent  UV ones we are looking for. This method has been suggested in 
\cite{Misiak:1994zw} and has been elaborated on in the context of three-loop
calculations in \cite{beta_den_comp}.
The resulting massive tadpole integrals can be computed with
the FORM-based program MATAD \cite{MATAD}. 

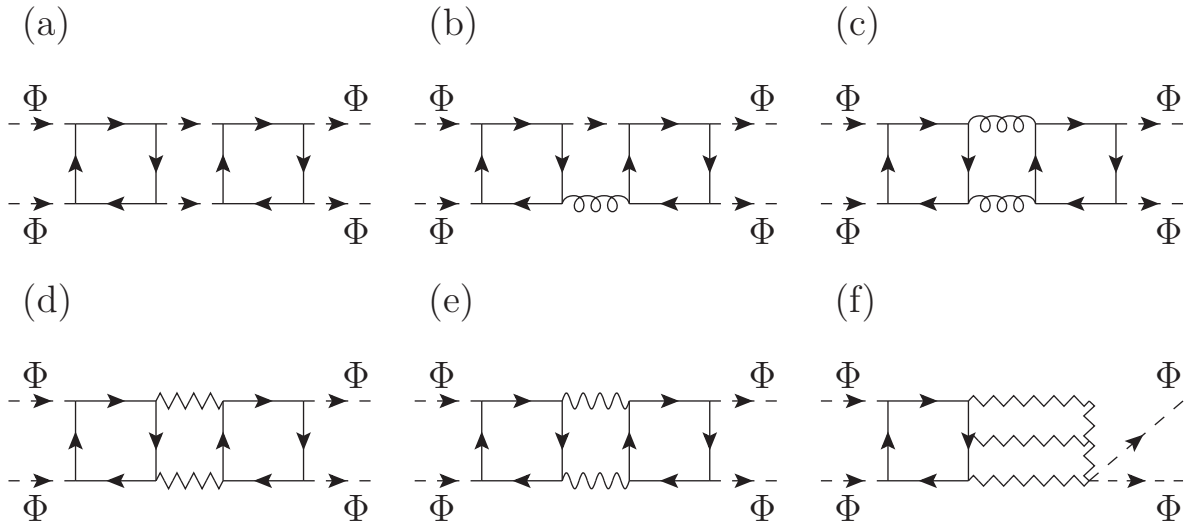
\begin{figure}[h!]
  \begin{tabular}{lll}
    \begin{picture}(140,100) (0,0)
    \SetWidth{0.5}
    \SetColor{Black}
    \DashArrowLine(0,15)(25,15){4}
    \DashArrowLine(0,45)(25,45){4}
    \DashArrowLine(110,15)(135,15){4}
    \DashArrowLine(110,45)(135,45){4}
    \ArrowLine(25,45)(55,45)
    \ArrowLine(55,15)(25,15)
    \ArrowLine(25,15)(25,45)
    \ArrowLine(55,45)(55,15)
    \DashArrowLine(55,15)(80,15){4}
    \DashArrowLine(55,45)(80,45){4}
    \ArrowLine(80,45)(110,45)
    \ArrowLine(110,15)(80,15)
    \ArrowLine(80,15)(80,45)
    \ArrowLine(110,45)(110,15)
    \Text(5,0)[lb]{\Large{\Black{$\Phi$}}}
    \Text(125,0)[lb]{\Large{\Black{$\Phi$}}}
    \Text(5,50)[lb]{\Large{\Black{$\Phi$}}}
    \Text(125,50)[lb]{\Large{\Black{$\Phi$}}}
    \Text(5,75)[lb]{\Large{\Black{(a)}}}
  \end{picture}
&    \begin{picture}(140,100) (0,0)
    \SetWidth{0.5}
    \SetColor{Black}
    \DashArrowLine(0,15)(25,15){4}
    \DashArrowLine(0,45)(25,45){4}
    \DashArrowLine(110,15)(135,15){4}
    \DashArrowLine(110,45)(135,45){4}
    \ArrowLine(25,45)(55,45)
    \ArrowLine(55,15)(25,15)
    \ArrowLine(25,15)(25,45)
    \ArrowLine(55,45)(55,15)
    \Gluon(55,15)(80,15){3}{3}
    \DashArrowLine(55,45)(80,45){4}
    \ArrowLine(80,45)(110,45)
    \ArrowLine(110,15)(80,15)
    \ArrowLine(80,15)(80,45)
    \ArrowLine(110,45)(110,15)
    \Text(5,0)[lb]{\Large{\Black{$\Phi$}}}
    \Text(125,0)[lb]{\Large{\Black{$\Phi$}}}
    \Text(5,50)[lb]{\Large{\Black{$\Phi$}}}
    \Text(125,50)[lb]{\Large{\Black{$\Phi$}}}
    \Text(5,75)[lb]{\Large{\Black{(b)}}}
  \end{picture}
&    \begin{picture}(140,100) (0,0)
    \SetWidth{0.5}
    \SetColor{Black}
    \DashArrowLine(0,15)(25,15){4}
    \DashArrowLine(0,45)(25,45){4}
    \DashArrowLine(110,15)(135,15){4}
    \DashArrowLine(110,45)(135,45){4}
    \ArrowLine(25,45)(55,45)
    \ArrowLine(55,15)(25,15)
    \ArrowLine(25,15)(25,45)
    \ArrowLine(55,45)(55,15)
    \Gluon(55,15)(80,15){3}{3}
    \Gluon(55,45)(80,45){3}{3}
    \ArrowLine(80,45)(110,45)
    \ArrowLine(110,15)(80,15)
    \ArrowLine(80,15)(80,45)
    \ArrowLine(110,45)(110,15)
    \Text(5,0)[lb]{\Large{\Black{$\Phi$}}}
    \Text(125,0)[lb]{\Large{\Black{$\Phi$}}}
    \Text(5,50)[lb]{\Large{\Black{$\Phi$}}}
    \Text(125,50)[lb]{\Large{\Black{$\Phi$}}}
    \Text(5,75)[lb]{\Large{\Black{(c)}}}
  \end{picture}\\
    \begin{picture}(140,100) (0,0)
    \SetWidth{0.5}
    \SetColor{Black}
    \DashArrowLine(0,15)(25,15){4}
    \DashArrowLine(0,45)(25,45){4}
    \DashArrowLine(110,15)(135,15){4}
    \DashArrowLine(110,45)(135,45){4}
    \ArrowLine(25,45)(55,45)
    \ArrowLine(55,15)(25,15)
    \ArrowLine(25,15)(25,45)
    \ArrowLine(55,45)(55,15)
    \ZigZag(55,15)(80,15){3}{4}
    \ZigZag(55,45)(80,45){3}{4}
    \ArrowLine(80,45)(110,45)
    \ArrowLine(110,15)(80,15)
    \ArrowLine(80,15)(80,45)
    \ArrowLine(110,45)(110,15)
    \Text(5,0)[lb]{\Large{\Black{$\Phi$}}}
    \Text(125,0)[lb]{\Large{\Black{$\Phi$}}}
    \Text(5,50)[lb]{\Large{\Black{$\Phi$}}}
    \Text(125,50)[lb]{\Large{\Black{$\Phi$}}}
    \Text(5,75)[lb]{\Large{\Black{(d)}}}
  \end{picture}
&    \begin{picture}(140,100) (0,0)
    \SetWidth{0.5}
    \SetColor{Black}
    \DashArrowLine(0,15)(25,15){4}
    \DashArrowLine(0,45)(25,45){4}
    \DashArrowLine(110,15)(135,15){4}
    \DashArrowLine(110,45)(135,45){4}
    \ArrowLine(25,45)(55,45)
    \ArrowLine(55,15)(25,15)
    \ArrowLine(25,15)(25,45)
    \ArrowLine(55,45)(55,15)
    \Photon(55,15)(80,15){3}{4}
    \Photon(55,45)(80,45){3}{4}
    \ArrowLine(80,45)(110,45)
    \ArrowLine(110,15)(80,15)
    \ArrowLine(80,15)(80,45)
    \ArrowLine(110,45)(110,15)
    \Text(5,0)[lb]{\Large{\Black{$\Phi$}}}
    \Text(125,0)[lb]{\Large{\Black{$\Phi$}}}
    \Text(5,50)[lb]{\Large{\Black{$\Phi$}}}
    \Text(125,50)[lb]{\Large{\Black{$\Phi$}}}
    \Text(5,75)[lb]{\Large{\Black{(e)}}}
  \end{picture}
&    \begin{picture}(140,100) (0,0)
    \SetWidth{0.5}
    \SetColor{Black}
    \DashArrowLine(0,15)(25,15){4}
    \DashArrowLine(0,45)(25,45){4}
    \DashArrowLine(100,15)(135,15){4}
    \DashArrowLine(100,15)(135,45){5}
    \ArrowLine(25,45)(55,45)
    \ArrowLine(55,15)(25,15)
    \ArrowLine(25,15)(25,45)
    \ArrowLine(55,45)(55,15)
    \ZigZag(55,15)(100,15){2}{6}
    \ZigZag(55,30)(100,30){2}{6}
    \ZigZag(55,45)(100,45){2}{6}
    \ZigZag(100,15)(100,45){2}{4}
    \Text(5,0)[lb]{\Large{\Black{$\Phi$}}}
    \Text(125,0)[lb]{\Large{\Black{$\Phi$}}}
    \Text(5,50)[lb]{\Large{\Black{$\Phi$}}}
    \Text(125,50)[lb]{\Large{\Black{$\Phi$}}}
    \Text(5,75)[lb]{\Large{\Black{(f)}}}
  \end{picture}
\end{tabular}
\caption{Some diagrams contributing to the renormalization of the $4$-$\Phi$-vertices}
\label{diasZ4ph}
\end{figure}
As opposed to the case of the Yukawa coupling $\beta$-functions (see \cite{Chetyrkin:2012rz,Bednyakov:2012en})
a completely naive treatment of $\gamma_5$ in dimensional regularization is possible for $\beta_\lambda$ and $\beta_{m^2}$.
In four dimensions we define
\be \gamma_5=i\gamma^0\gamma^1\gamma^2\gamma^3=\f{i}{4!}\eps_{\mu\nu\rho\sigma} \gamma^\mu\gamma^\nu\gamma^\rho\gamma^\sigma \text{ with }
 \eps_{0123}=1=-\eps^{0123} \label{gamma5} 
{}.
\ee
In order to have a non-naive contribution from a fermion loop with a $\gamma_5$ matrix in it at least 
four free Lorentz indices or momenta on the external lines of the minimal subgraph containing this fermion loop are required.
These can be indices from the gauge boson vertices or the internal momenta from other loops which act as external momenta to 
the minimal subgraph containing the fermion loop in question. 
External momenta of the whole diagram can be set to zero as the renormalization constants in the $\overline{\text{MS}}$-scheme do not
depend on those.
In four dimensions the trace of such a fermion line
will produce a result $\propto \eps_{\mu_1\mu_2\mu_3\mu_4}$ where $\mu_1, \ldots,\mu_4$ are the aforementioned
free Lorentz indices.
If there is a second fermion line the trace of which also yields an $\eps$-tensor these two $\eps$-tensors can be contracted
and we can get a non-naive contribution from $\gamma_5$.

Let us consider two examples. The diagram in Fig.\ref{diasZ4ph} (c) has two fermion loops. If we take one of the fermion loops 
with the momenta on the two external $\Phi$-legs set to zero we have two indices from the
gluon lines attached to the fermion loop and one loop momentum going through the two gluons and acting as an external momentum
to the subgraph containing only this fermion loop. This is not enough to have a non-naive $\gamma_5$ contribution from this graph.

The fermion loop in Fig.\ref{diasZ4ph} (f) has three Lorentz indices and two external loop momenta but there is
no second fermion line to produce a second $\eps$-tensor. Furthermore - as we set all external momenta to zero - there are no free 
Lorentz indices or momenta in the final result to support an $\eps$-tensor there. Therefore any contribution with an $\eps$-tensor 
from this
fermion loop (which is the only antisymmetric Lorentz structure) must vanish after contraction with the Lorentz structures from
the W-bosons.

\section{Results \label{res:beta}}
In this section we give the results for the three-loop $\beta$-functions for the 
couplings $\lambda$ and the mass parameter $m^2$ setting all gauge group factors to their SM values.
All results of this work can found at\\
\texttt{\bf http://www-ttp.particle.uni-karlsruhe.de/Progdata/ttp13/ttp13-008/}.\footnote{
There we also present the results in a form where the QCD colour factors are not set to numbers
but generically expressed through the quadratic Casimir operators $\cf$ and $\ca$ of the 
quark  and the adjoint representation of the corresponding Lie algebra,
the dimension of the quark representation $\dR$ and the trace
$\tr$ defined through \mbox{$\tr \delta^{ab}=\textbf{Tr}\lb T^a T^b\rb$}  
with the group generators $T^a$ of the quark representation.}\\
We denote the number of generations by $\NGen$.

\be
\begin{split}
\beta_{\sss{\lambda}}^{(1)}=&   
       - \ytau^4    
       - \yb^4    3     
       + \gw^4    \f{9}{16}   
       + \gb^2 \gw^2    \f{3}{8}     
       + \gb^4   \f{3}{16}    
       + \lambda \ytau^2    2    
       + \lambda \yb^2     6         
       - \lambda \gw^2     \f{9}{2}         
       - \lambda \gb^2   \f{3}{2}         \\ &   
       + \lambda^2     12        
       + \yt^2 \lambda    6         
       - \yt^4    3         \,,\\
\beta_{\sss{\lambda}}^{(2)}=&     
        \ytau^6     5         
       + \yb^6         15      
       - \gw^4 \ytau^2     \f{3}{8}    
       - \gw^4 \yb^2    \f{9}{8}      
       + \gw^6    \left(          \f{497}{32}          - 2 \NGen          \right)    
       - \gb^2 \ytau^4   2         
       + \gb^2 \yb^4     \f{2}{3}              \\ &   
       + \gb^2 \gw^2 \ytau^2     \f{11}{4} 
       + \gb^2 \gw^2 \yb^2    \f{9}{4}  
       - \gb^2 \gw^4    \left(           \f{97}{96}          + \f{2}{3} \NGen      \right)
       - \gb^4 \ytau^2   \f{25}{8}    
       + \gb^4 \yb^2   \f{5}{8}            \\ &  
       - \gb^4 \gw^2    \left(          \f{239}{96}          + \f{10}{9} \NGen      \right)    
       - \gb^6    \left(           \f{59}{96}          + \f{10}{9} \NGen       \right)
       - \lambda \ytau^4     \f{1}{2}         
       - \lambda \yb^4    \f{3}{2}           \\ &    
       + \lambda \gw^2 \ytau^2   \f{15}{4}        
       + \lambda \gw^2 \yb^2     \f{45}{4}  
       + \lambda \gw^4    \left(          - \f{313}{16}          + 5 \NGen          \right)    
       + \lambda \gb^2 \ytau^2     \f{25}{4}    
       + \lambda \gb^2 \yb^2     \f{25}{12}          \\ &     
       + \lambda \gb^2 \gw^2     \f{39}{8}      
       + \lambda \gb^4    \left(           \f{229}{48}      + \f{25}{9} \NGen    \right)
       - \lambda^2 \ytau^2    24        
       - \lambda^2 \yb^2    72       
       + \lambda^2 \gw^2   54           \\ &      
       + \lambda^2 \gb^2     18     
       - \lambda^3     156           
       - \yt^2 \yb^4  3      
       - \yt^2 \gw^4     \f{9}{8}   
       + \yt^2 \gb^2 \gw^2    \f{21}{4}  
       - \yt^2 \gb^4    \f{19}{8}          \\ &     
       - \yt^2 \lambda \yb^2     21       
       + \yt^2 \lambda \gw^2    \f{45}{4}         
       + \yt^2 \lambda \gb^2   \f{85}{12}        
       - \yt^2 \lambda^2     72         
       - \yt^4 \yb^2     3       
       - \yt^4 \gb^2   \f{4}{3}         \\ &     
       - \yt^4 \lambda   \f{3}{2}   
       + \yt^6    15     
       - \gs^2 \yb^4   16            
       + \gs^2 \lambda \yb^2     40         
       + \gs^2 \yt^2 \lambda     40         
       - \gs^2 \yt^4   16   .\\ 
\end{split} \label{beta:la1l2l}
\ee

\begin{align}
\beta_{\sss{\lambda}}^{(3)}=&
   \ytau^8    \left(
          - \f{143}{8}
          - 12 \zeta_{3}
          \right)
       - \yb^2 \ytau^6    
           \f{297}{8}         
       - \yb^4 \ytau^4    72   
       - \yb^6 \ytau^2     \f{297}{8}
       + \yb^8    \left(
          - \f{1599}{8}
          - 36 \zeta_{3}
          \right)                 \notag \\ &  
       + \gw^2 \ytau^6    \left(
           \f{1137}{32}
          - 9 \zeta_{3}
          \right)
       + \gw^2 \yb^6    \left(
           \f{3411}{32}
          - 27 \zeta_{3}
          \right)
       + \gw^4 \ytau^4    \left(
           \f{4503}{128}
          - \f{273}{16} \zeta_{3}
          - \f{13}{4} \NGen
          \right)         \notag \\ &  
       + \gw^4 \yb^2 \ytau^2   \f{9}{8}
       + \gw^4 \yb^4    \left(
           \f{13653}{128}
          - \f{819}{16} \zeta_{3}
          - \f{39}{4} \NGen
          \right)
       + \gw^6 \ytau^2    \left(
          - \f{5739}{256}
          + \f{99}{4} \zeta_{3}
          + \f{9}{2} \NGen
          \right)         \notag \\ &  
       + \gw^6 \yb^2    \left(
          - \f{17217}{256}
          + \f{297}{4} \zeta_{3}
          + \f{27}{2} \NGen
          \right)
       + \gw^8    \left(
          \f{982291}{3072}
          - \f{2781}{128} \zeta_{3}
          - \f{14749}{192} \NGen \right.  \notag \\ & \left.
          - 45 \NGen \zeta_{3}
          - \f{5}{3} \NGen^2
          \right)         
       + \gb^2 \ytau^6    \left(
          \f{135}{32}
          + 33 \zeta_{3}
          \right)
       + \gb^2 \yb^6    \left(
          \f{5111}{96}
          - 25 \zeta_{3}
          \right)                      \notag \\ &
       + \gb^2 \gw^2 \ytau^4    \left(
          - \f{15}{64}
          - \f{381}{8} \zeta_{3}
          \right)
       - \gb^2 \gw^2 \yb^2 \ytau^2      \f{5}{4}
       + \gb^2 \gw^2 \yb^4    \left(
          - \f{3239}{192}
          - \f{311}{8} \zeta_{3}
          \right)              \notag \\ &
       + \gb^2 \gw^4 \ytau^2    \left(
           \f{1833}{256}
          - \f{3}{2} \zeta_{3}
          - \f{1}{2} \NGen
          \right)  
       + \gb^2 \gw^4 \yb^2    \left(
           \f{4179}{256}
          + 9 \zeta_{3}
          + \f{5}{2} \NGen
          \right)             \notag \\ &
       + \gb^2 \gw^6    \left(
          - \f{54053}{3456}
          - \f{405}{32} \zeta_{3}
          - \f{8341}{864} \NGen
          - \f{10}{27} \NGen^2
          \right)
       + \gb^4 \ytau^4    \left(
           \f{5697}{128}
          + \f{375}{16} \zeta_{3}
          + \f{65}{12} \NGen
          \right)            \notag \\ &
       + \gb^4 \yb^2 \ytau^2  \f{41}{24}
         + \gb^4 \yb^4    \left(
           \f{15137}{3456}
          - \f{2035}{144} \zeta_{3}
          - \f{415}{36} \NGen
          \right)
       + \gb^4 \gw^2 \ytau^2    \left(
          \f{6657}{256}
          - \f{15}{2} \zeta_{3}
          - \f{5}{6} \NGen
          \right)            \notag \\ &
       + \gb^4 \gw^2 \yb^2    \left(
           \f{4403}{256}
          + \f{9}{2} \zeta_{3}
          + \f{25}{6} \NGen
          \right)
       + \gb^4 \gw^4    \left(
          - \f{64693}{3456}
          + \f{873}{64} \zeta_{3}
          + \f{149}{648} \NGen
          + 7 \NGen \zeta_{3} \right.  \notag \\ & \left.
          - \f{50}{81} \NGen^2
          \right)           
       + \gb^6 \ytau^2    \left(
           \f{3929}{256}
          - \f{15}{4} \zeta_{3}
          + \f{55}{6} \NGen
          \right)
       + \gb^6 \yb^2    \left(
           \f{12043}{2304}
          + \f{5}{4} \zeta_{3}
          + \f{95}{18} \NGen
          \right)           \notag \\ &
       + \gb^6 \gw^2    \left(
          - \f{29779}{6912}
          + \f{75}{32} \zeta_{3}
          - \f{18001}{2592} \NGen
          + \f{61}{9} \NGen \zeta_{3}
          - \f{250}{243} \NGen^2
          \right)           \notag \\ &
       + \gb^8    \left(
          - \f{6845}{9216}
          + \f{99}{128} \zeta_{3}
          - \f{20735}{1728} \NGen
          + \f{95}{9} \NGen \zeta_{3}
          - \f{125}{81} \NGen^2
          \right)  
       + \lambda \ytau^6    \left(
          - \f{1241}{8}
          - 66 \zeta_{3}
          \right)         \notag \\ &
       + \lambda \yb^2 \ytau^4  240
       + \lambda \yb^4 \ytau^2    240
       + \lambda \yb^6    \left(
           \f{117}{8}
          - 198 \zeta_{3}
          \right)  
       + \lambda \gw^2 \ytau^4    \left(
          - \f{1587}{8}
          + 171 \zeta_{3}
          \right)        \notag \\ &
       - \lambda \gw^2 \yb^2 \ytau^2   27
       + \lambda \gw^2 \yb^4    \left(
          - \f{4977}{8}
          + 513 \zeta_{3}
          \right)  
       + \lambda \gw^4 \ytau^2    \left(
          - \f{1311}{64}
          - \f{117}{2} \zeta_{3}
          - \f{21}{4} \NGen
          \right)        \notag \\ &
       + \lambda \gw^4 \yb^2    \left(
          - \f{3933}{64}
          - \f{351}{2} \zeta_{3}
          - \f{63}{4} \NGen
          \right)
       + \lambda \gw^6    \left(
          - \f{46489}{288}
          + \f{2259}{8} \zeta_{3}
          + \f{3515}{36} \NGen  \right.   \notag \\ &\left.
          + 90 \NGen \zeta_{3}  
          + \f{70}{9} \NGen^2
          \right)
       + \lambda \gb^2 \ytau^4    \left(
           \f{507}{8}
          - 117 \zeta_{3}
          \right)
       - \lambda \gb^2 \yb^2 \ytau^2 9  
       + \lambda \gb^2 \yb^4    \left(
          - \f{5737}{24}    \right.   \notag \\ &\left.
          + 249 \zeta_{3}
          \right)
       + \lambda \gb^2 \gw^2 \ytau^2    \left(
          - \f{3771}{32}
          + 126 \zeta_{3}
          \right)
       + \lambda \gb^2 \gw^2 \yb^2    \left(
          - \f{3009}{32}
          + 12 \zeta_{3}
          \right)  \notag \\ &
       + \lambda \gb^2 \gw^4    \left(
           \f{4553}{32}
          - \f{249}{8} \zeta_{3}
          + \f{33}{2} \NGen
          - 6 \NGen \zeta_{3}
          \right)
       + \lambda \gb^4 \ytau^2    \left(
          - \f{1783}{64}
          - \f{123}{2} \zeta_{3}
          - \f{65}{4} \NGen
          \right) \notag \\ &
       + \lambda \gb^4 \yb^2    \left(
          - \f{127303}{1728}
          - \f{47}{6} \zeta_{3}
          - \f{155}{36} \NGen
          \right)
       + \lambda \gb^4 \gw^2    \left(
           \f{979}{8}
          - \f{3}{8} \zeta_{3}
          + \f{95}{4} \NGen
          - 6 \NGen \zeta_{3}
          \right)\notag \\ &
       + \lambda \gb^6    \left(
           \f{12679}{432}
          + \f{9}{8} \zeta_{3}
          + \f{5995}{162} \NGen
          - \f{190}{9} \NGen \zeta_{3}
          + \f{1750}{243} \NGen^2
          \right)
       + \lambda^2 \ytau^4    \left(
           \f{717}{2}
          + 252 \zeta_{3}
          \right)\notag \\ &
       - \lambda^2 \yb^2 \ytau^2    216
       + \lambda^2 \yb^4    \left(
           \f{1719}{2}
          + 756 \zeta_{3}
          \right)
       + \lambda^2 \gw^2 \ytau^2    \left(
           \f{213}{4}
          - 144 \zeta_{3}
          \right)           
       + \lambda^2 \gw^2 \yb^2    \left(
           \f{639}{4}    \right.   \notag \\ &\left.
          - 432 \zeta_{3}
          \right) 
       + \lambda^2 \gw^4    \left(
           \f{1995}{8}
          - 513 \zeta_{3}
          - 141 \NGen
          \right)
       + \lambda^2 \gb^2 \ytau^2    \left(
          - \f{541}{4}
          + 96 \zeta_{3}
          \right)           \notag \\ &
       + \lambda^2 \gb^2 \yb^2    \left(
           \f{417}{4}
          - 192 \zeta_{3}
          \right) 
       + \lambda^2 \gb^2 \gw^2    \left(
          - 333
          - 162 \zeta_{3}
          \right)
       + \lambda^2 \gb^4    \left(
          - 183
          - 81 \zeta_{3} \right.   \notag \\ &\left.
          - \f{235}{3} \NGen
          \right)    
       + \lambda^3 \ytau^2   291
       + \lambda^3 \yb^2   873 
       + \lambda^3 \gw^2    \left(
          - 474
          + 72 \zeta_{3}
          \right)
       + \lambda^3 \gb^2    \left(
          - 158
          + 24 \zeta_{3}
          \right)         \notag \\ &
       + \lambda^4    \left(
           3588
          + 2016 \zeta_{3}
          \right)     
       + \yt^2 \ytau^6    \left(
          - \f{297}{8}
          \right)
       + \yt^2 \yb^2 \ytau^4   12
       + \yt^2 \yb^4 \ytau^2   \f{45}{8} 
       + \yt^2 \yb^6    \left(
          - \f{717}{8} \right.   \notag \\ &\left.
          - 36 \zeta_{3}
          \right)
       + \yt^2 \gw^2 \yb^4    \f{477}{32}
       + \yt^2 \gw^4 \ytau^2   \f{9}{8}
       + \yt^2 \gw^4 \yb^2    \left(
          - \f{351}{64}
          + \f{117}{2} \zeta_{3}
          - 12 \NGen
          \right)       \notag \\ &
       + \yt^2 \gw^6    \left(
          - \f{17217}{256}
          + \f{297}{4} \zeta_{3}
          + \f{27}{2} \NGen
          \right)
       + \yt^2 \gb^2 \yb^4    \left(
          - \f{2299}{96}
          + 26 \zeta_{3}
          \right)      \notag \\ &
       + \yt^2 \gb^2 \gw^2 \ytau^2   \f{29}{4}
       + \yt^2 \gb^2 \gw^2 \yb^2    \left(
           \f{1001}{96}
          + \f{31}{2} \zeta_{3}
          \right)
       + \yt^2 \gb^2 \gw^4    \left(
           \f{3103}{256}
          + \f{27}{4} \zeta_{3}
          + \f{1}{2} \NGen
          \right)      \notag \\ &
       + \yt^2 \gb^4 \ytau^2  \f{701}{24}
       + \yt^2 \gb^4 \yb^2    \left(
          - \f{709}{64}
          - \zeta_{3}
          \right)
       + \yt^2 \gb^4 \gw^2    \left(
           \f{23521}{768}
          - 3 \zeta_{3}
          + \f{5}{6} \NGen
          \right)      \notag \\ &
       + \yt^2 \gb^6    \left(
           \f{42943}{2304}
          - \f{5}{2} \zeta_{3}
          + \f{215}{18} \NGen
          \right)
       + \yt^2 \lambda \ytau^4   240
       + \yt^2 \lambda \yb^2 \ytau^2    21     \notag \\ &
       + \yt^2 \lambda \yb^4    \left(
           \f{6399}{8}
          + 144 \zeta_{3}
          \right)
       - \yt^2 \lambda \gw^2 \ytau^2    27
       + \yt^2 \lambda \gw^2 \yb^2    \left(
          - \f{531}{4}
          + 54 \zeta_{3}
          \right)   \notag \\ &
       + \yt^2 \lambda \gw^4    \left(
          - \f{3933}{64}
          - \f{351}{2} \zeta_{3}
          - \f{63}{4} \NGen
          \right)
       - \yt^2 \lambda \gb^2 \ytau^2    9
       + \yt^2 \lambda \gb^2 \yb^2    \left(
          - \f{929}{12}
          - 2 \zeta_{3}
          \right)   \notag \\ &
       + \yt^2 \lambda \gb^2 \gw^2    \left(
          - \f{6509}{32}
          + 177 \zeta_{3}
          \right)
       + \yt^2 \lambda \gb^4    \left(
          - \f{112447}{1728}
          - \f{449}{6} \zeta_{3}
          - \f{635}{36} \NGen
          \right)  \notag \\ &
       - \yt^2 \lambda^2 \ytau^2    216
       + \yt^2 \lambda^2 \yb^2    \left(
           117
          - 864 \zeta_{3}
          \right)
       + \yt^2 \lambda^2 \gw^2    \left(
           \f{639}{4}
          - 432 \zeta_{3}
          \right)                \notag \\ &
       + \yt^2 \lambda^2 \gb^2    \left(
          - \f{195}{4}
          - 48 \zeta_{3}
          \right)
       + \yt^2 \lambda^3    873
       - \yt^4 \ytau^4    72
       + \yt^4 \yb^2 \ytau^2     \f{45}{8}
       + \yt^4 \yb^4    72 \zeta_{3}               \notag \\ &
       + \yt^4 \gw^2 \yb^2    \f{477}{32}
       + \yt^4 \gw^4    \left(
           \f{13653}{128}
          - \f{819}{16} \zeta_{3}
          - \f{39}{4} \NGen
          \right)
       + \yt^4 \gb^2 \yb^2    \left(
           \f{1337}{96}
          - 28 \zeta_{3}
          \right)               \notag \\ &
       + \yt^4 \gb^2 \gw^2    \left(
          - \f{1079}{192}
          - \f{743}{8} \zeta_{3}
          \right)
       + \yt^4 \gb^4    \left(
           \f{100913}{3456}
          + \f{2957}{144} \zeta_{3}
          - \f{115}{36} \NGen
          \right)               \notag \\ &
       + \yt^4 \lambda \ytau^2  240
       + \yt^4 \lambda \yb^2    \left(
           \f{6399}{8}
          + 144 \zeta_{3}
          \right)
       + \yt^4 \lambda \gw^2    \left(
          - \f{4977}{8}
          + 513 \zeta_{3}
          \right)              \notag \\ &
       + \yt^4 \lambda \gb^2    \left(
          - \f{2485}{24}
          + 57 \zeta_{3}
          \right)
       + \yt^4 \lambda^2    \left(
           \f{1719}{2}
          + 756 \zeta_{3}
          \right)
       - \yt^6 \ytau^2   \f{297}{8}            \notag \\ &
       + \yt^6 \yb^2    \left(
          - \f{717}{8}
          - 36 \zeta_{3}
          \right)  
       + \yt^6 \gw^2    \left(
           \f{3411}{32}
          - 27 \zeta_{3}
          \right)
       + \yt^6 \gb^2    \left(
           \f{3467}{96}
          + 17 \zeta_{3}
          \right)            \notag \\ &
       + \yt^6 \lambda    \left(
           \f{117}{8}
          - 198 \zeta_{3}
          \right)
       + \yt^8    \left(
          - \f{1599}{8}
          - 36 \zeta_{3}
          \right)
       + \gs^2 \yb^6    \left(
          - 38
          + 240 \zeta_{3}
          \right)         \notag \\ &
       + \gs^2 \gw^2 \yb^4    \left(
          - \f{31}{2}
          + 24 \zeta_{3}
          \right)
       + \gs^2 \gw^4 \yb^2    \left(
           \f{651}{8}
          - 54 \zeta_{3}
          \right)
       + \gs^2 \gw^6    \left(
          - \f{153}{8} \NGen
          + 18 \NGen \zeta_{3}
          \right)       \notag \\ &
       + \gs^2 \gb^2 \yb^4    \left(
          - \f{641}{18}
          + \f{136}{3} \zeta_{3}
          \right)
       + \gs^2 \gb^2 \gw^2 \yb^2    \left(
           \f{233}{4}
          - 36 \zeta_{3}
          \right)
       + \gs^2 \gb^2 \gw^4    \left(
          - \f{51}{8} \NGen
          + 6 \NGen \zeta_{3}
          \right)      \notag \\ &
       + \gs^2 \gb^4 \yb^2    \left(
          \f{683}{24}
          - 18 \zeta_{3}
          \right)
       + \gs^2 \gb^4 \gw^2    \left(
          - \f{187}{24} \NGen
          + \f{22}{3} \NGen \zeta_{3}
          \right)
       + \gs^2 \gb^6    \left(
          - \f{187}{24} \NGen   \right. \notag \\ & \left. 
          + \f{22}{3} \NGen \zeta_{3}
          \right)     
       + \gs^2 \lambda \yb^4    \left(
          895
          - 1296 \zeta_{3}
          \right)
       + \gs^2 \lambda \gw^2 \yb^2    \left(
          - \f{489}{2}
          + 216 \zeta_{3}
          \right) \notag \\ & 
       + \gs^2 \lambda \gw^4    \left(
           \f{135}{2} \NGen
          - 72 \NGen \zeta_{3}
          \right)
       + \gs^2 \lambda \gb^2 \yb^2    \left(
          - \f{991}{18}
          + 40 \zeta_{3}
          \right)
       + \gs^2 \lambda \gb^4    \left(
           \f{55}{2} \NGen  \right. \notag \\ & \left. 
          - \f{88}{3} \NGen \zeta_{3}
          \right)
       + \gs^2 \lambda^2 \yb^2    \left(
          - 1224
          + 1152 \zeta_{3}
          \right)
       + \gs^2 \yt^2 \yb^4    \left(
          - 2
          - 48 \zeta_{3}
          \right)         \notag \\ &
       + \gs^2 \yt^2 \gw^2 \yb^2    \left(
          - 8
          + 96 \zeta_{3}
          \right)
       + \gs^2 \yt^2 \gw^4    \left(
           \f{651}{8}
          - 54 \zeta_{3}
          \right)
       + \gs^2 \yt^2 \gb^2 \gw^2    \left(
           \f{249}{4}
          - 36 \zeta_{3}
          \right)       \notag \\ &
       + \gs^2 \yt^2 \gb^4    \left(
           \f{587}{24}
          - 18 \zeta_{3}
          \right)
       + \gs^2 \yt^2 \lambda \yb^2    \left(
           82
          - 96 \zeta_{3}
          \right)
       + \gs^2 \yt^2 \lambda \gw^2    \left(
          - \f{489}{2}
          + 216 \zeta_{3}
          \right)       \notag \\ &
       + \gs^2 \yt^2 \lambda \gb^2    \left(
          - \f{2419}{18}
          + 136 \zeta_{3}
          \right)
       + \gs^2 \yt^2 \lambda^2    \left(
          - 1224
          + 1152 \zeta_{3}
          \right)
       + \gs^2 \yt^4 \yb^2    \left(
          - 2
          - 48 \zeta_{3}
          \right)     \notag \\ &
       + \gs^2 \yt^4 \gw^2    \left(
          - \f{31}{2}
          + 24 \zeta_{3}
          \right)
       + \gs^2 \yt^4 \gb^2    \left(
           \f{931}{18}
          - \f{56}{3} \zeta_{3}
          \right)
       + \gs^2 \yt^4 \lambda    \left(
           895
          - 1296 \zeta_{3}
          \right)  \notag \\ &
       + \gs^2 \yt^6    \left(
          - 38
          + 240 \zeta_{3}
          \right)
       + \gs^4 \yb^4    \left(
          - \f{626}{3}
          + 32 \zeta_{3}
          + 40 \NGen
          \right)
       + \gs^4 \lambda \yb^2    \left(
           \f{1820}{3}    \right. \notag \\ & \left. 
          - 48 \zeta_{3}
          - 64 \NGen
          \right)
       + \gs^4 \yt^2 \yb^2   192
       + \gs^4 \yt^2 \lambda    \left(
           \f{1820}{3}
          - 48 \zeta_{3}
          - 64 \NGen
          \right)\notag \\ &
       + \gs^4 \yt^4    \left(
          - \f{626}{3}
          + 32 \zeta_{3}
          + 40 \NGen
          \right){}.
\label{beta:la3l}      
\end{align} 
The purely $\lambda$-dependent part of eq.~(\ref{beta:la3l}) has been derived before in  
\cite{Brezin:1974xi,Brezin:1973}, the full one-loop
and two-loop results in eq.~(\ref{beta:la1l2l}) are in agreement with \cite{2loopbetayukawa,Machacek198570,Ford:1992pn} and in the limit
$g_1,g_2,\yb,\ytau\rightarrow 0$ we reproduce the result \cite{Chetyrkin:2012rz} at three-loop level.

The running of the $m^2$ parameter is given by
\be
\begin{split}
\f{\beta_{\sss{m^2}}^{(1)}}{m^2}=& 
              \ytau^2 
       + \yb^2  3
       - \gw^2   \f{9}{4}
        - \gb^2  \f{3}{4}
               + \lambda  6
         + \yt^2  3{},\\
\f{\beta_{\sss{m^2}}^{(2)}}{m^2}=&  
            - \ytau^4  \f{9}{4}        
       - \yb^4  \f{27}{4}       
       + \gw^2 \ytau^2  \f{15}{8}        
       + \gw^2 \yb^2  \f{45}{8}     \\&  
       + \gw^4   \left(          - \f{385}{32}          + \f{5}{2} \NGen          \right)
       + \gb^2 \ytau^2 \f{25}{8}          
       + \gb^2 \yb^2  \f{25}{24}          
       + \gb^2 \gw^2  \f{15}{16}     \\&    
       + \gb^4    \left(          + \f{157}{96}          + \f{25}{18} \NGen          \right)
       - \lambda \ytau^2         12        
       - \lambda \yb^2            36   
       + \lambda \gw^2            36        
       + \lambda \gb^2            12  \\&        
       - \lambda^2                30        
       - \yt^2 \yb^2    \f{21}{2}     
       + \yt^2 \gw^2     \f{45}{8}    
       + \yt^2 \gb^2    \f{85}{24}       
       - \yt^2 \lambda   36     \\&      
       - \yt^4   \f{27}{4}   
       + \gs^2 \yb^2  20         
       + \gs^2 \yt^2  20             ,\\ 
       \end{split}
\ee
 \be      \begin{split}
\f{\beta_{\sss{m^2}}^{(3)}}{m^2}=& 
  \ytau^6    \left(
          - \f{233}{16}
          + 15 \zeta_{3}
          \right)
       + \yb^2 \ytau^4    72
       + \yb^4 \ytau^2     72
       + \yb^6    \left(
           \f{1605}{16}
          + 45 \zeta_{3}
          \right)                 \\&  
       + \gw^2 \ytau^4    \left(
          - \f{987}{16}
          + 54 \zeta_{3}
          \right)         
       - \gw^2 \yb^2 \ytau^2     \f{27}{2}
       + \gw^2 \yb^4    \left(
          - \f{3177}{16}
          + 162 \zeta_{3}
          \right)              \\&  
       + \gw^4 \ytau^2    \left(
          - \f{255}{128}
          - \f{81}{4} \zeta_{3}
          - \f{21}{8} \NGen
          \right)                
       + \gw^4 \yb^2    \left(
          - \f{765}{128}
          - \f{243}{4} \zeta_{3}
          - \f{63}{8} \NGen
          \right)            \\&  
       + \gw^6    \left(
          - \f{39415}{576}
          + \f{711}{16} \zeta_{3}
          + \f{2867}{72} \NGen
          + 45 \NGen \zeta_{3}
          + \f{35}{9} \NGen^2
          \right)                       
       + \gb^2 \ytau^4    \left(
           \f{291}{16}
          - 36 \zeta_{3}
          \right)              \\&  
       - \gb^2 \yb^2 \ytau^2    \f{9}{2}
       + \gb^2 \yb^4    \left(
          - \f{1067}{16}
          + 72 \zeta_{3}
          \right)                             
       + \gb^2 \gw^2 \ytau^2    \left(
          - \f{2331}{64}
          + 45 \zeta_{3}
          \right)                     
       - \gb^2 \gw^2 \yb^2    \f{865}{64}     \\&
       + \gb^2 \gw^4    \left(
           \f{2691}{64}
          - \f{405}{16} \zeta_{3}
          + \f{21}{4} \NGen
          - 3 \NGen \zeta_{3}
          \right)          
       + \gb^4 \ytau^2    \left(
          - \f{3607}{128}
          - \f{15}{4} \zeta_{3}
          - \f{65}{8} \NGen
          \right)                        \\&  
       + \gb^4 \yb^2    \left(
          - \f{79207}{3456}
          - \f{35}{12} \zeta_{3}
          - \f{155}{72} \NGen
          \right)                             
       + \gb^4 \gw^2    \left(
          \f{1053}{32}
          - \f{207}{16} \zeta_{3}
          + \f{55}{8} \NGen
          - 3 \NGen \zeta_{3}
          \right)                               \\& 
       + \gb^6    \left(
           \f{839}{108}
          - \f{51}{16} \zeta_{3}
          + \f{4375}{324} \NGen
          - \f{95}{9} \NGen \zeta_{3}
          + \f{875}{243} \NGen^2
          \right)                
       + \lambda \ytau^4    \left(
           \f{261}{4}
          + 72 \zeta_{3}
          \right)                               \\&  
       - \lambda \yb^2 \ytau^2   108
       + \lambda \yb^4    \left(
           \f{351}{4}
          + 216 \zeta_{3}
          \right)                        
       + \lambda \gw^2 \ytau^2    \left(
           \f{189}{8}
          - 108 \zeta_{3}
          \right)
       + \lambda \gw^2 \yb^2    \left(
           \f{567}{8}
          - 324 \zeta_{3}
          \right)                          \\&   
       + \lambda \gw^4    \left(
           \f{11511}{32}
          - 162 \zeta_{3}
          - \f{153}{2} \NGen
          \right)                       
       + \lambda \gb^2 \ytau^2    \left(
          - \f{549}{8}
          + 36 \zeta_{3}
          \right)
       + \lambda \gb^2 \yb^2    \left(
           \f{393}{8}
          - 132 \zeta_{3}
          \right)                    \\&  
       + \lambda \gb^2 \gw^2    \left(
          - \f{1701}{16}
          + 36 \zeta_{3}
          \right)                            
       + \lambda \gb^4    \left(
          - \f{1077}{32}
          - 18 \zeta_{3}
          - \f{85}{2} \NGen
          \right)
       + \lambda^2 \ytau^2    \f{99}{2}
       + \lambda^2 \yb^2    \f{297}{2}                        \\&  
       + \lambda^2 \gw^2    \left(
          - 63
          - 108 \zeta_{3}
          \right)
       + \lambda^2 \gb^2    \left(
          - 21
          - 36 \zeta_{3}
          \right)
       + \lambda^3    1026
       + \yt^2 \ytau^4    72                       
       + \yt^2 \yb^2 \ytau^2    \f{21}{2}  \\&  
       + \yt^2 \yb^4    \left(
           \f{4047}{16}
          + 36 \zeta_{3}
          \right)
       + \yt^2 \gw^2 \ytau^2    \left(
          - \f{27}{2}
          \right)
       + \yt^2 \gw^2 \yb^2    \left(
          - \f{243}{8}
          - 27 \zeta_{3}
          \right)                            \\&  
       + \yt^2 \gw^4    \left(
          - \f{765}{128}
          - \f{243}{4} \zeta_{3}
          - \f{63}{8} \NGen
          \right)
       - \yt^2 \gb^2 \ytau^2   \f{9}{2}
       + \yt^2 \gb^2 \yb^2    \left(
          - \f{929}{24}
          - \zeta_{3}
          \right)                            \\&  
       + \yt^2 \gb^2 \gw^2    \left(
          - \f{3277}{64}
          + \f{117}{2} \zeta_{3}
          \right)
       + \yt^2 \gb^4    \left(
          - \f{123103}{3456}
          - \f{149}{12} \zeta_{3}
          - \f{635}{72} \NGen
          \right)
       - \yt^2 \lambda \ytau^2    108                           \\&  
       + \yt^2 \lambda \yb^2    \left(
          - \f{315}{2}
          - 216 \zeta_{3}
          \right)
       + \yt^2 \lambda \gw^2    \left(
           \f{567}{8}
          - 324 \zeta_{3}
          \right)
       + \yt^2 \lambda \gb^2    \left(
          - \f{219}{8}
          - 60 \zeta_{3}
          \right)                            \\&  
       + \yt^2 \lambda^2     \f{297}{2}
       + \yt^4 \ytau^2     72
       + \yt^4 \yb^2    \left(
          \f{4047}{16}
          + 36 \zeta_{3}
          \right)
       + \yt^4 \gw^2    \left(
          - \f{3177}{16}
          + 162 \zeta_{3}
          \right)                            \\&  
       + \yt^4 \gb^2    \left(
          - \f{431}{16}
          + 12 \zeta_{3}
          \right)
       + \yt^4 \lambda    \left(
           \f{351}{4}
          + 216 \zeta_{3}
          \right)
       + \yt^6    \left(
          \f{1605}{16}
          + 45 \zeta_{3}
          \right)                            \\&  
       + \gs^2 \yb^4    \left(
           \f{447}{2}
          - 360 \zeta_{3}
          \right)
       + \gs^2 \gw^2 \yb^2    \left(
          - \f{489}{4}
          + 108 \zeta_{3}
          \right)
       + \gs^2 \gw^4    \left(
           \f{135}{4} \NGen
          - 36 \NGen \zeta_{3}
          \right)                            \\&  
       + \gs^2 \gb^2 \yb^2    \left(
          - \f{991}{36}
          + 20 \zeta_{3}
          \right)
       + \gs^2 \gb^4    \left(
           \f{55}{4} \NGen
          - \f{44}{3} \NGen \zeta_{3}
          \right)
       + \gs^2 \lambda \yb^2    \left(
          - 612
          + 576 \zeta_{3}
          \right)                                  \\&  
       + \gs^2 \yt^2 \yb^2    \left(
           41
          - 48 \zeta_{3}
          \right)                         
       + \gs^2 \yt^2 \gw^2    \left(
          - \f{489}{4}
          + 108 \zeta_{3}
          \right)
       + \gs^2 \yt^2 \gb^2    \left(
          - \f{2419}{36}
          + 68 \zeta_{3}
          \right)                         \\&                
       + \gs^2 \yt^2 \lambda    \left(
          - 612
          + 576 \zeta_{3}
          \right)                 
       + \gs^2 \yt^4    \left(
           \f{447}{2}
          - 360 \zeta_{3}
          \right)        \\&        
       + \gs^4 \yb^2    \left(
           \f{910}{3}
          - 24 \zeta_{3}
          - 32 \NGen
          \right)
       + \gs^4 \yt^2    \left(
           \f{910}{3}
          - 24 \zeta_{3}
          - 32 \NGen
          \right){}.\\
\end{split}
\ee
The one-loop and two-loop parts of this result are in agreement with
\cite{2loopbetayukawa,Ford:1992pn}. The purely $\lambda$-dependent part can be found in \cite{Brezin:1974xi,Brezin:1973}
and for $g_1,g_2,\yb,\ytau\rightarrow 0$ we reproduce the result \cite{Chetyrkin:2012rz} again.

Now we want to give a numerical evaluation of the $\beta$-functions at the scale of the top mass in order to get an idea
of the size of the new terms.
For \mbox{$M_t \approx 173.5$ GeV}, \mbox{$M_H \approx 126$ GeV} and \mbox{$\als=0.1184$} \cite{pdg2012}
we get the couplings in the $\overline{\text{MS}}$-scheme at this scale  using one-loop matching relations
\cite{Espinosa:2007qp,Hempfling:1994ar,Sirlin1986389}:\footnote{This has been done with the Mathematica package
SMPoleMatching.m by F. Bezrukov which can be downloaded at \texttt{http://www.inr.ac.ru/\~{}fedor/SM/download.php}}
\mbox{$\yt(M_t)\approx 0.94$}, \mbox{$\gs(M_t)\approx 1.1644$}, \mbox{$\gw(M_t)\approx 0.6484$}, \mbox{$\gb(M_t)\approx 0.3587$}
and \mbox{$\lambda(M_t)\approx 0.13$}. The lighter Yukawa couplings can be estimated from the $\overline{\text{MS}}$-masses
$m_b \approx 4.18$ GeV and $m_\tau \approx 1.777$ GeV \cite{pdg2012} to be
\mbox{$\yb\approx\sqrt{2}\f{m_b}{v}\approx 0.02$} and 
\mbox{$\ytau\approx\sqrt{2}\f{m_\tau}{v}\approx 0.01$}

For $\beta_{\lambda}(\mu=M_t)$ we find at one-loop order
\be\begin{split}
\f{\beta_{\sss{\lambda}}^{(1)}(\mu=M_t)}{(16\pi^2)}=&
\underbrace{-1.5\times 10^{-2}}_{ \yt^4}
\underbrace{+4.4\times 10^{-3}}_{ \lambda \yt^2}
\underbrace{-1.6\times 10^{-3}}_{ \gw^2 \lambda}
\underbrace{+1.3\times 10^{-3}}_{ \lambda^2}\\ &
\underbrace{+6.3\times 10^{-4}}_{ \gw^4}
\underbrace{-1.6\times 10^{-4}}_{ \lambda \gb^2}
\underbrace{+1.3\times 10^{-4}}_{ \gw^2 \gb^2}
\underbrace{2.\times 10^{-5}}_{ \gb^4}\\ &
\underbrace{+2.\times 10^{-6}}_{ \lambda \yb^2}
\underbrace{+1.6\times 10^{-7}}_{ \lambda \ytau^2}
\underbrace{-3.\times 10^{-9}}_{ \yb^4}
\underbrace{-6.3\times 10^{-11}}_{ \ytau^4}\,.
\end{split}
\ee

At two-loop order the largest terms with $\yb$ are of $\mathcal{O}(10^{-7})$
and those with $\ytau$ of $\mathcal{O}(10^{-9})$. Neglecting these small terms
we find

\be\begin{split}
\f{\beta_{\sss{\lambda}}^{(2)}(\mu=M_t)}{(16\pi^2)^2}=&
\underbrace{-6.8\times 10^{-4} }_{ \gs^2 \yt^4}
\underbrace{+4.1\times 10^{-4} }_{ \yt^6}
\underbrace{+2.5\times 10^{-4} }_{ \gs^2 \lambda \yt^2}
\underbrace{-4.3\times 10^{-5} }_{ \lambda^2 \yt^2}\\ &
\underbrace{+2.8\times 10^{-5} }_{ \gw^6}
\underbrace{+2.2\times 10^{-5} }_{ \gw^2 \lambda \yt^2}
\underbrace{+1.5\times 10^{-5} }_{ \gw^2 \lambda^2}
\underbrace{-1.4\times 10^{-5} }_{ \lambda^3}\\ &
\underbrace{+1.\times 10^{-5} }_{ \gw^2 \yt^2 \gb^2}
\underbrace{-7.\times 10^{-6} }_{ \gw^4 \yt^2}
\underbrace{-6.1\times 10^{-6} }_{ \lambda \yt^4}
\underbrace{-5.4\times 10^{-6} }_{ \yt^4 \gb^2}\\ &
\underbrace{-4.2\times 10^{-6} }_{ \gw^4 \lambda}
\underbrace{+4.2\times 10^{-6} }_{ \lambda \yt^2 \gb^2}
\underbrace{-2.7\times 10^{-6} }_{ \gw^4 \gb^2}
\underbrace{-1.6\times 10^{-6} }_{ \gw^2 \gb^4}\\ &
\underbrace{+1.6\times 10^{-6} }_{ \lambda^2 \gb^2}
\underbrace{-1.4\times 10^{-6} }_{ \yt^2 \gb^4}
\underbrace{+1.4\times 10^{-6} }_{ \gw^2 \lambda \gb^2}
\underbrace{+1.1\times 10^{-6} }_{ \lambda \gb^4}\\ &
\underbrace{-3.4\times 10^{-7} }_{ \gb^6 }+\text{(terms $\propto \yb^2,\ytau^2$)}\,.
\end{split}
\ee
At three-loop level we only give the largest terms and omit small ones 
of $\mathcal{O}(10^{-7})$:

\be\begin{split}
\f{\beta_{\sss{\lambda}}^{(3)}(\mu=M_t)}{(16\pi^2)^3}=&
\underbrace{5.9\times 10^{-5}}_{  \gs^2 \yt^6}
\underbrace{-3.8\times 10^{-5}}_{  \yt^8}
\underbrace{-2.3\times 10^{-5}}_{  \gs^2 \lambda \yt^4}
\underbrace{+1.9\times 10^{-5}}_{  \gs^4 \lambda \yt^2}\\ &
\underbrace{-1.8\times 10^{-5}}_{  \gs^4 \yt^4}
\underbrace{+5.9\times 10^{-6}}_{  \lambda^2 \yt^4}
\underbrace{+5.5\times 10^{-6}}_{  \gw^2 \yt^6}
\underbrace{-5.1\times 10^{-6}}_{  \lambda \yt^6}\\ &
\underbrace{+2.1\times 10^{-6} }_{ \gw^6 \lambda}
\underbrace{-1.6\times 10^{-6} }_{ \gw^4 \lambda \yt^2}
\underbrace{+1.5\times 10^{-6} }_{ \gw^2 \gs^2 \yt^4}
\underbrace{+1.3\times 10^{-6} }_{ \yt^6 \gb^2}\\ &
\underbrace{-1.3\times 10^{-6} }_{ \gw^2 \yt^4 \gb^2}
\underbrace{+1.\times 10^{-6} }_{ \gw^6 \yt^2}
\underbrace{+1.\times 10^{-6} }_{ \gs^2 \yt^4 \gb^2}
\underbrace{-9.1\times 10^{-7} }_{ \gw^8}\\ &
\underbrace{+8.9\times 10^{-7} }_{ \gw^4 \gs^2 \yt^2}
\underbrace{+8.3\times 10^{-7} }_{ \gs^2 \lambda^2 \yt^2}
\underbrace{-6.\times 10^{-7} }_{ \gw^4 \lambda^2}
\underbrace{+5.6\times 10^{-7} }_{ \gw^4 \yt^4}\\ &
\underbrace{-4.5\times 10^{-7} }_{ \gw^4 \gs^2 \lambda}
\underbrace{+4.4\times 10^{-7} }_{ \lambda^4}
\underbrace{+4.3\times 10^{-7} }_{ \lambda^3 \yt^2}
+\text{smaller terms}\,.
\end{split}
\ee

The dominant contributions contain only $\gs,\yt$ and $\lambda$ as suggested in \cite{Chetyrkin:2012rz}.
This result therefore explicitly confirms the validity of the approximation made in \cite{Chetyrkin:2012rz} at
the scale of the top mass for the individual terms of the $\beta$-function.
On the other hand, the cancellations between some of these individual terms are huge. Especially the terms containing only $\gs,\yt$ and $\lambda$
cancel so well at the scale of the top mass that the overall value of the three-loop $\beta$-function at this scale
\mbox{$\f{\beta_{\sss{\lambda}}^{(3)}(\mu=M_t)}{(16\pi^2)^3}=1.1 \times 10^{-5}$} is about a factor $5$ larger than with the electroweak
contributions neglected \mbox{($\left.\f{\beta_{\sss{\lambda}}^{(3)}(\mu=M_t)}{(16\pi^2)^3}\right|_{\gw,\gb\rightarrow 0}=2.1 \times 10^{-6}$)}.

\begin{figure}[h!]
\includegraphics[width=\textwidth]{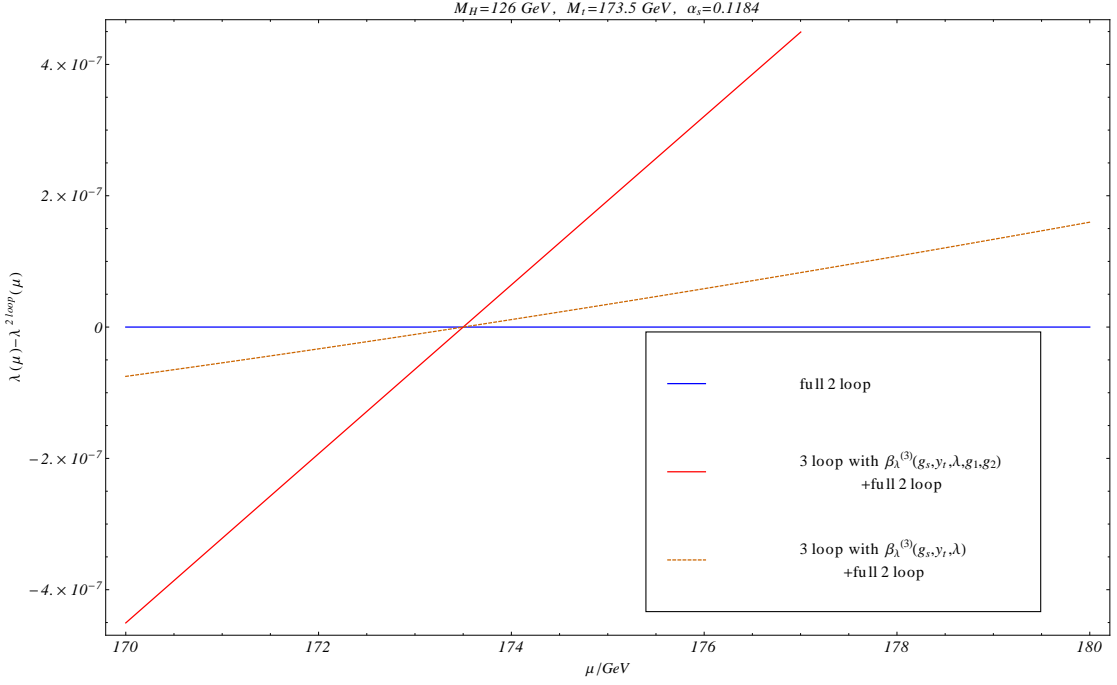} 
\caption{Evolution of $\lambda$ around the scale of the top mass the with and without the
electroweak three-loop contributions.}
\label{lambda170GeV_diff}
\end{figure}

\begin{figure}[h!]
\includegraphics[width=\textwidth]{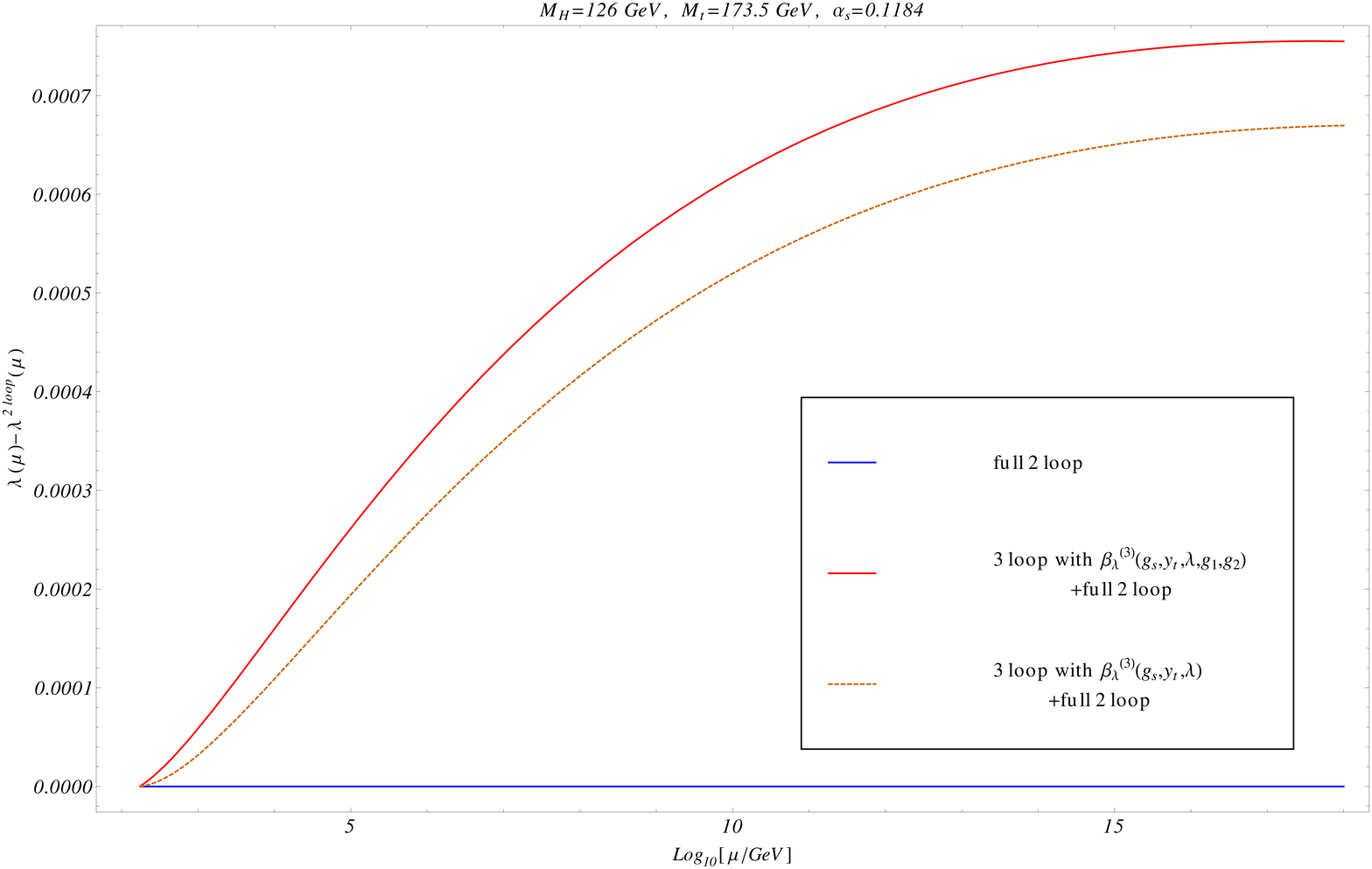} 
\caption{Evolution of $\lambda$ around the scale of the top mass the with and without the
electroweak three-loop contributions.}
\label{lambda10to2_18GeV_diff}
\end{figure}

\begin{figure}[h!]
\includegraphics[width=\textwidth]{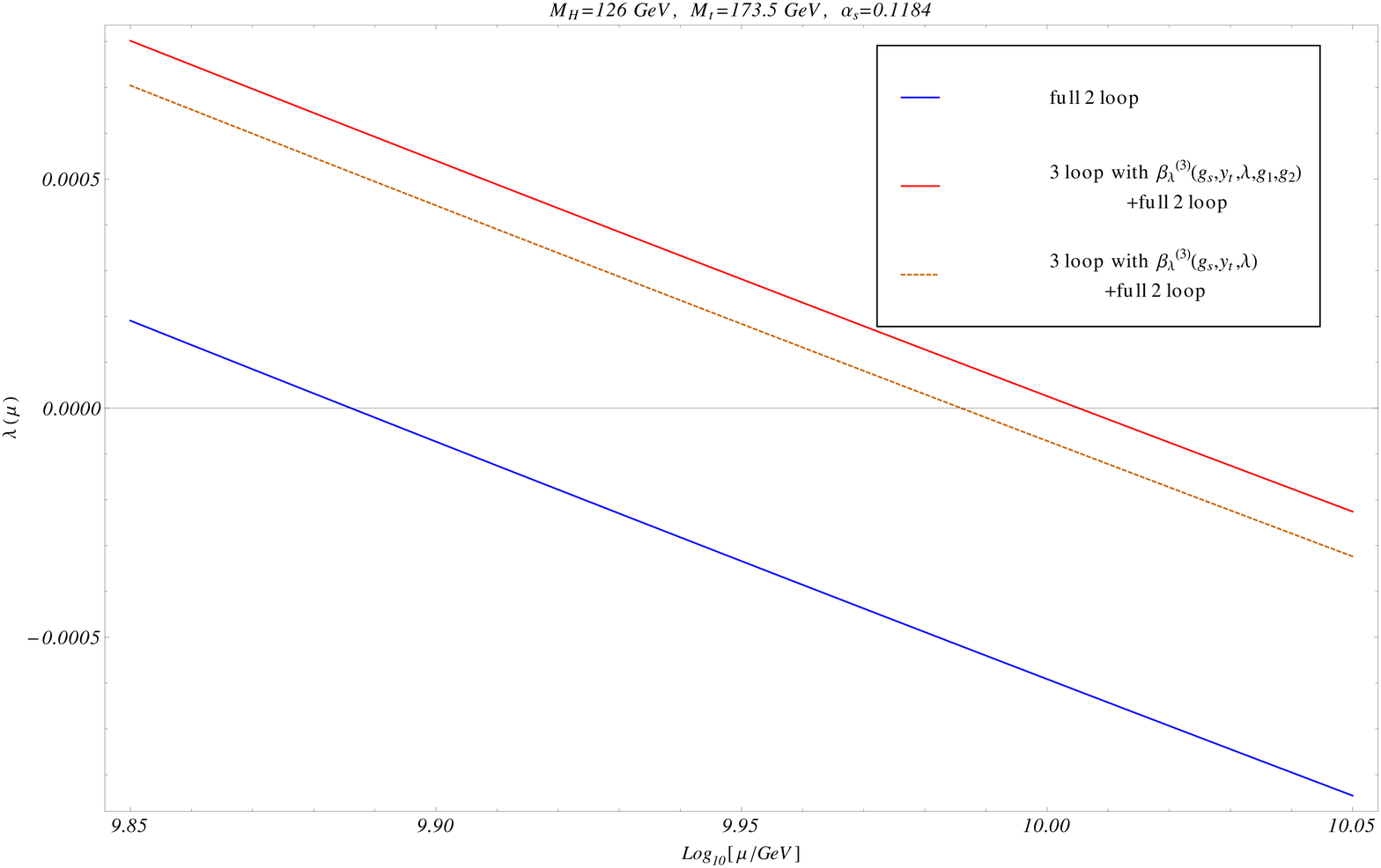} 
\caption{Evolution of $\lambda$ with the scale $\mu$ around $10^{10}$ GeV with and without the
electroweak three-loop contributions.}
\label{lambda10to10GeV}
\end{figure}

Their effect on the running of $\lambda$, however, is not as strong at higher scales. 
In Fig. \ref{lambda170GeV_diff} the difference $\lambda-\lambda^{2 loop}$ is shown around the scale of the top mass. $\lambda^{2 loop}$ is the Higgs self-interaction evolved with
the full two-loop function $\beta_\lambda$ and $\lambda$ is evolved once including all three-loop contributions and once including only $\gs, \yt$ and $\lambda$, neglecting the electroweak couplings.
At one and two-loop level the electroweak contributions are always included and the $\beta$-functions for all other couplings are taken at three-loop level including all couplings $\gs,\gw,\gb, \yt$ and $\lambda$.
As starting values for the couplings we use the ones given above.
Here we see that the gradient of the full three-loop curve is about a factor of $5$ larger than the one for the curve without the electroweak three-loop corrections,
which is in agreement with the numerics presented above.
If we plot the difference $\lambda-\lambda^{2 loop}$ from the top mass scale up to the Planck scale, which is shown in Fig. \ref{lambda10to2_18GeV_diff},
we see that the difference between the curves for $\beta_\lambda$ with and without the three-loop electroweak corrections grows strongest at low scales and stays almost constant at higher scales,
where the impact of the terms with $g_1$ and $g_2$ decreases. This means that for the evolution of $\lambda$ up to high scales the terms with only $\gs, \yt$ and $\lambda$ are indeed the most
important ones as assumed in \cite{Chetyrkin:2012rz}. Nevertheless, at low scales the new contributions presented in this paper should be included due to huge cancellations among the $\gs, \yt$ and $\lambda$ terms.
The effect of the electroweak terms at low scales carries of course to the large scales which makes them important for a precision analysis of questions like SM vacuum stability.
To illustrate this we plot in Fig. \ref{lambda10to10GeV} the evolution of $\lambda$ at the scale where $\lambda$ becomes negative.\footnote{Note that there are huge experimental errors on the input parameters, 
especially the top mass, which make an accurate prediction of where this transition happens (or if at all) impossible for the moment. The relative positions of the curves in this plot, however,
illustrate nicely the effect of higher orders in the $\beta$-function for $\lambda$. For more details see e.g. \cite{Chetyrkin:2012rz}.}

The whole $\beta$-function is dominated by the negative one-loop term $\propto \yt^4$ but with the new terms presented here
$\lambda$ will decrease less which means that these contributions slightly
enhance the stability of the electroweak vacuum state in the SM.

For $\beta_{m^2}(\mu=M_t)$ we find at one-loop order
\be\begin{split}
\f{\beta_{\sss{m^2}}^{(1)}(\mu=M_t)}{(16\pi^2) m^2}=&
\underbrace{1.7\times10^{-2}}_{ \yt^2}
\underbrace{-6.\times10^{-3} }_{\gw^2}
\underbrace{+4.9\times10^{-3}}_{ \lambda}
\underbrace{-6.1\times10^{-4}}_{ \gb^2}\\ &
\underbrace{+7.6\times10^{-6}}_{ \yb^2}
\underbrace{+6.3\times10^{-7}}_{ \ytau^2}\,.
\end{split}
\ee

Omitting the small contributions proportional to $\yb$ and $\ytau$ 
the two-loop terms look like this:
\be\begin{split}
\f{\beta_{\sss{m^2}}^{(2)}(\mu=M_t)}{(16\pi^2)^2 m^2}=&
\underbrace{9.6\times10^{-4}}_{ \gs^{2} \yt^{2}}
\underbrace{-2.1\times10^{-4}}_{ \yt^{4}}
\underbrace{-1.7\times10^{-4}}_{ \lambda \yt^{2}}\\ &
\underbrace{+8.4\times10^{-5}}_{ \gw^{2} \yt^{2}}
\underbrace{+7.9\times10^{-5}}_{ \gw^{2} \lambda}
\underbrace{-3.2\times10^{-5}}_{ \gw^{4}}
\underbrace{-2.\times10^{-5}}_{ \lambda^{2}}
\underbrace{+1.6\times10^{-5}}_{ \yt^{2} \gb^{2}}\\ &
\underbrace{+8.\times10^{-6} }_{\lambda \gb^{2}}
\underbrace{+3.9\times10^{-6}}_{\gb^{4}}
\underbrace{+2.\times10^{-6} }_{\gw^{2} \gb^{2}}+\text{(terms $\propto \yb^2,\ytau^2$)}\,.
\end{split}
\ee
Again we only give the largest terms at three-loop level and omit small ones 
of $\mathcal{O}(10^{-7})$:
\be\begin{split}
\f{\beta_{\sss{m^2}}^{(3)}(\mu=M_t)}{(16\pi^2)^3 m^2}=&
\underbrace{+7.4\times10^{-5} }_{ \gs^4 \yt^2}
\underbrace{-5.6\times10^{-5} }_{ \gs^2 \yt^4}
\underbrace{+2.7\times10^{-5} }_{ \yt^6}
\underbrace{+9.\times10^{-6} }_{ \lambda \yt^4}\\&
\underbrace{+5.7\times10^{-6} }_{ \gw^6}
\underbrace{-4.1\times10^{-6}  }_{\gw^4 \yt^2}
\underbrace{-3.9\times10^{-6} }_{ \gw^2 \lambda \yt^2}
\underbrace{+3.2\times10^{-6} }_{ \gs^2 \lambda \yt^2}
\underbrace{-1.7\times10^{-6} }_{ \gw^4 \gs^2}\\&
\underbrace{+9.7\times10^{-7} }_{ \gw^2 \gs^2 \yt^2}
\underbrace{+5.7\times10^{-7} }_{ \lambda^3}
\underbrace{+5.7\times10^{-7} }_{ \gs^2 \yt^2 \gb^2}
\underbrace{+5.6\times10^{-7} }_{ \lambda^2 \yt^2}+\text{smaller terms}\,.
\end{split}
\ee
We see that especially the term $\propto \gw^6$ is not much smaller than
the three dominant terms that have already been computed in \cite{Chetyrkin:2012rz}.
The overall three-loop results with $\gw,\gb\rightarrow 0$ and the electroweak
interaction switched on differ only by $5.\times10^{-6}$ due to cancellations
between the new terms.

\section{Conclusions \label{last}}

We have computed the three-loop $\beta$-functions for the
quartic Higgs self-coupling and for the mass parameter $m^2$ in the unbroken phase of the SM,
neglecting only the Yukawa couplings of the first two generations and the mixing of quark generations
(an extension that could easily be made but which is numerically negligible).

The electroweak contributions are small as expected which confirms the validity of the
approximation made in our previous calculation \cite{Chetyrkin:2012rz}.
The impact of these new terms is strongest at the scale of the top mass due to huge cancellations
among the QCD, Yukawa top and Higgs self-interaction terms. At higher scales they become negligible.
Nevertheless, as the couplings are measured at low scales and evolved fom there, including
the electroweak contributions in a precision analysis of the evolution of $\lambda$ is
important.

{\it Note added:} Recently, a similar calculation has independently confirmed the results presented in this paper \cite{Bednyakov:2013eba}.

\section*{Acknowledgements}
We thank  Johann K\"uhn for useful discussions and support. We would also like to thank A. Bednyakov for useful correspondence
which led to the finding of inaccuracies in the numerics of the first version of this work.
 
Finally we want to mention that all our calculations have been
performed using the thread-based \cite{Tentyukov:2007mu} version  of FORM
\cite{Vermaseren:2000nd}.  The Feynman diagrams  have been drawn with the 
Latex package Axodraw \cite{Vermaseren:1994je}.

This work has been supported by the Deutsche Forschungsgemeinschaft in the
Sonderforschungsbereich/Transregio SFB/TR-9 ``Computational Particle
Physics''. M.Z. has been supported by the Graduiertenkolleg ``Elementarteilchenphysik
bei h\"ochsten Energien und h\"ochster Pr\"azission''.

\bibliographystyle{JHEP}

\bibliography{LiteraturSM}

\end{document}

%% file: macros.tex
\usepackage[latin1]{inputenc} 
\usepackage[T1]{fontenc}
\usepackage[safe]{textcomp}
\usepackage{lmodern} 
\usepackage{epsfig}
\usepackage{float}
\usepackage{amstext}  
\usepackage{amsfonts}
\usepackage{amssymb} 
\usepackage{amsmath}
\usepackage{amsthm}
\usepackage{bm}       
\usepackage[bottom]{footmisc} 
\usepackage{array}                
\usepackage{xcolor} 
\usepackage{framed}
\usepackage{slashed}
\usepackage[absolute]{textpos}
\usepackage{axodraw4j}
\usepackage{dsfont}
\usepackage{multirow}

\usepackage{appendix}
\usepackage{listings}
\usepackage{enumerate}     
\usepackage[bottom]{footmisc} 
\usepackage{array}           
\usepackage{parskip}        
\usepackage{xcolor} 
\usepackage{color}
\usepackage{framed}
\usepackage{fancyhdr}
\usepackage{subfig}
\usepackage[raggedright]{sidecap}

\usepackage{tikz}
\usepackage[absolute]{textpos}
\usepackage{pstricks}
\usepackage{axodraw4j}
\usepackage{graphicx}

\newcommand{\p}{\partial}

\newcommand{\f}[2]{\frac{#1}{#2}}
\newcommand{\sss}[1]{\scriptscriptstyle#1}
\newcommand{\vv}[2]{\left( \begin{array}{c} #1 \\ #2  \end{array} \right)}
\newcommand{\bea}{\begin{eqnarray}}
\newcommand{\eea}{\end{eqnarray}}
\newcommand{\be}{\begin{equation}}
\newcommand{\ee}{\end{equation}}
\newcommand{\ba}{\begin{align}}
\newcommand{\ea}{\end{align}}
\newcommand{\beas}{\begin{eqnarray*}}
\newcommand{\eeas}{\end{eqnarray*}}
\newcommand{\bes}{\begin{equation*}}
\newcommand{\ees}{\end{equation*}}
\newcommand{\bas}{\begin{align*}}
\newcommand{\eas}{\end{align*}}

\newcommand{\ssL}{{\mathcal L}} 
\newcommand{\eps}{{\varepsilon}}
 
\newcommand{\cf}{C_{\scriptscriptstyle{F}}} 
\newcommand{\ca}{C_{\scriptscriptstyle{A}}}
\newcommand{\tr}{T_{\scriptscriptstyle{F}}}

\newcommand{\dR}{d_{\scriptscriptstyle{R}}}

\newcommand{\NGen}{N_{\scriptscriptstyle{g}}}

\newcommand{\gs}{g_{\scriptscriptstyle{s}}}
\newcommand{\yt}{y_{\scriptscriptstyle{t}}}
\newcommand{\yb}{y_{\scriptscriptstyle{b}}}
\newcommand{\ytau}{y_{\scriptscriptstyle{\tau}}}
\newcommand{\gb}{g_1}
\newcommand{\gw}{g_2}

\newcommand{\als}{\alpha_{\scriptscriptstyle{s}}}

\newcommand{\lb}{\left(}
\newcommand{\rb}{\right)}

\definecolor{bluemar}{rgb}{0,0,.5}
\definecolor{redmar}{rgb}{.8,0,0}
\definecolor{greenmar}{rgb}{0,.5,0}